\documentclass[twocolumn,showpacs,preprintnumbers,amsmath,amssymb]{revtex4}
\usepackage{graphicx}% Include figure files
\usepackage{dcolumn}% Align table columns on decimal point
\usepackage{bm}% bold math

\newcounter{sub}
\newcounter{subeqn}[sub]
\def\be{\begin{equation}}
\def\ee{\end{equation}}

\def\lp{\left(}
\def\rp{\right)}

\def\ls{\left[}
\def\rs{\right]}

\def\st{\stepcounter{sub}}
\def\stq{\stepcounter{subeqn}}
\def\bea{\begin{eqnarray}}
\def\eea{\end{eqnarray}}

\newcommand\xxi{{\mbox{\boldmath $\xi$}}}

\newcommand\nab{\mbox{\boldmath $\nabla$}}

\def\b{{\bf b}}

\def\v{{\bf v}}
\def\u{{\bf u}}
\def\e{{\bf e}}
\def\r{{\bf r}}

\newcommand\B{{\bf B}}

\newcommand\p{{\bf p}}

\newcommand\no{\nonumber}

\begin{document}

\preprint{}

\title{QPOs and firehose instabilities in neutron star magnetospheres in accreting systems }
% Force line breaks with \\

\author{Vahid Rezania}
 \altaffiliation[Also at ]{Institute for Advanced Studies in Basic Sciences,
          Zanjan 45195, Iran.}
\email{vrezania@phys.ualberta.ca}
\homepage{http://fermi.phys.ualberta.ca/~vrezania}
 %Lines break automatically or can be forced with \\
\author{J. C. Samson}%

\affiliation{%
    Department of Physics,
    University of Alberta\\
    Edmonton, AB, Canada, T6G 2J1
}%

\date{\today}% It is always \today, today,
             %  but any date may be explicitly specified

\begin{abstract}
We show that the interaction of an accretion disk with the
magnetosphere of a neutron star can excite resonant shear Alfv\'en
waves with Hz-kHz frequencies in a region of enhanced density
gradients. This is the the region where accretion material flows
along the magnetic field lines in the magnetosphere. We argue that
due to the pressure anisotropy produced by the plasma flow, firehose
instabilities are likely to occur. Furthermore, for a dipolar field
topology, we show that a new instability develops due to both
magnetic field curvature and the plasma flow.

\end{abstract}

% PACS, the Physics and Astronomy
                             % Classification Scheme.
%\keywords{Suggested keywords}%Use showkeys class option if keyword
                              %display desired
\maketitle

%\section{\label{sec:int}Introduction}

Recent observations of quasi-periodic oscillations (QPOs),
particularly kHz QPOs, in X-ray emissions from accreting binaries,
have aroused a lot of interest in the astrophysical community. The
QPOs are very strong and remarkably coherent with frequencies
ranging from $\sim 10$ Hz to $\sim 1200$ Hz. They have been observed
in the X-ray flux of about 20 accreting neutron star sources and
five black hole sources by {\it Rossi X-Ray Timing Explorer}. Almost
all sources have shown twin spectral peaks in the QPOs in the kHz
part of the X-ray spectrum, with the value of the peak separation
being anti-correlated with the QPO frequencies
\cite{Van00,DMV03,MVF04}.

The clear similarities of kHz QPO properties in black hole systems
to those in neutron star binaries \cite{PBV99,BPV02}, and in white
dwarf systems \cite{WWP03}, suggest that the oscillation likely
originate in the accretion flow surrounding the central object.
Motivated by this argument, a variety of accretion-based models have
been proposed. Most models, however, fail to provide a general
explanation of QPO features in all potential sources. See, for
example, Li \& Narayan \cite{LN04} and references therein.

Li \& Narayan \cite{LN04} studied Rayleigh-Taylor and
Kelvin-Helmholtz instabilities at a possible interface between the
star's magnetosphere and the accretion disk. They found that modes
with low order azimuthal wavenumbers are expected to grow to large
amplitude and to contribute to kHz QPOs. In their study, the
magnetic field allows an interface with abrupt spatial
discontinuities in the flow density and/or angular velocity across
the interface. They ignore, however, the  dynamical role of the
magnetic field of the central object and its interaction with the
accretion disk/flow.

Based on theoretical models and observations of the aurora in the
Earth's magnetosphere, Rezania et al. \cite{RSD04} and Rezania \&
Samson \cite{RS05} (hereafter paper I) have recently proposed a
generic magnetospheric model for accretion disk-neutron star systems
to address the occurrence and the behavior of the observed QPOs in
those systems. In the Earth's magnetosphere, the occurrence of
aurora is a result of the resonant coupling between shear Alfv\'en
waves and fast compressional waves (produced by the solar wind).
These resonances are known as field line resonances (FLRs). Paper I
argued that this resonant coupling is also likely to occur in
neutron star magnetospheres, due to interaction with the accreting
plasma. The MHD interaction of the infalling plasma (with a
sonic/supersonic speed) with the neutron star magnetosphere, would
alter not only the plasma flow toward the surface of the star, as
assumed by current QPO models, but also the structure of the star's
magnetosphere.  The magnetic field of the neutron star is distorted
inward by the infalling plasma of the Keplerian accretion flow.
Furthermore, the plasma would likely be able to penetrate through
magnetic field lines and produce enhanced density regions within the
magnetosphere (similar to the magnetospheric interface introduced in
\cite{LN04}). Any instability in the compressional action of the
accretion flow would alter the quasi-equilibrium pressure balance
between the inward pressure of the infalling flow $\sim\rho v_r^2/2$
and the outward magnetic pressure $\sim B_p^2/8\pi$. This process
then can excite perturbations in the enhanced density region. Here
$\rho$ and $v_r$ are the density and radial velocity of the
infalling matter and $B_p$ is the poloidal magnetic field on the
plane of the disk. In analogy with the interaction of the solar wind
with the earth's magnetosphere \cite{Sam91}, one would expect the
excitation of resonant shear Alfv\'en waves, or FLRs, \cite{RS05}.

Paper I assumed a simple geometry with a rectilinear magnetic field,
and showed that, in the presence of a plasma flow,two resonant MHD
modes with frequencies in the kHz range can occur within a few
stellar radii. The results in paper I gave a reasonable prediction,
both quantitative and qualitative, of the kHz oscillations observed
in the X-ray fluxes in X-ray binaries. The existence of a non-zero
plasma displacement along the magnetic field lines, which oscillates
with resonance frequencies and then modulates the flow of the plasma
toward the surface of the neutron star, can explain the kHz
quasi-periodicity in the observed X-ray flux from the star. See
\cite{RS05} for more details.

In this letter we examine the above problem in a more realistic
configuration for the star's magnetosphere. We base are model on the
fact that the topology of the magnetospheres of isolated pulsars is
likely close to dipolar. In accreting pulsars, however, the geometry
of the magnetosphere will be distorted due to the inward flowing
plasma, specifically on the plane of the accretion disk.
Nevertheless, we approximate the topology of the magnetosphere of an
accreting neutron stars with a dipolar geometry in order to study
the structure of shear Alfv\'en waves in the presence of an ambient
flow.

Furthermore, due to the existence of ambient flow along the magnetic
field, the plasma pressure is not expected to be isotropic. This can
be understood by noting that the plasma pressure parallel to the
field lines can be defined as
\st \be\label{p_par}
p_{||}\sim m \int f (u_{||}-v_p)^2 d^3u,
\ee
where $u_{||}=\u\cdot\B/B$ is the
thermal velocity of the plasma particles and $v_p$ is the ambient
flow velocity along the field lines.  Here $f(\r,\u,t)$ is the
plasma distribution function satisfying the Vlasov equation
\cite{SHD97}. Similarly, the perpendicular component of the pressure
can be calculated from
\st \be\label{p_per} p_\perp \sim m \int f
u_\perp^2 d^3u/2.
\ee
Hence, the scalar pressure of isotropic MHD
must be replaced by a diagonal pressure tensor with two components:
a parallel component $p_{||}$ acting along the field lines and a
perpendicular component $p_\perp$ acting in the perpendicular
direction. The latter can be considered to be the ram pressure.
Consequently, in the MHD equations, the flow pressure must be
written as: ${\p}=p_\perp {\bf I} + (p_\perp-p_{||}) \b\b$, where
${\bf I}$ is the identity tensor and $\b=\B/B$ is the unit vector along
the magnetic field line.

The linearized, perturbed magnetohydrodynamic equations with
anisotropic pressure in the presence of an ambient flow now have
parallel and perpendicular components:
\st\begin{eqnarray}\label{eq-mhd1} \stq\label{xi_para} && \rho \lp
{\partial \delta \v \over  \partial t} +  \v \cdot \nab \delta\v +
\delta\v \cdot \nab \v\rp_{||} =
- \nabla_{||} \delta p_{||}   \no\\
&&\hspace{4cm}+ (p_\perp-p_{||}) {\nabla_{||}\delta B_{||}\over B}\\
\stq\label{xi_perp}
&& \rho \lp {\partial \delta \v \over  \partial t}
+  \v \cdot \nab
\delta\v +  \delta\v \cdot \nab \v\rp_\perp =
- \nab_\perp \delta \lp p_\perp + {B^2\over 8\pi}\rp\no\\
&&\hspace{3cm}
+\Xi \lp{ \delta\B \cdot \nab \B +  \B \cdot \delta\nab \B \over 4\pi}\rp_\perp\\
\stq\label{dB_t}
&&{\partial \delta \B\over\partial t} = \nab \times (\delta \v \times \B + \v
\times \delta \B), \\
\stq\label{div_dB}
&&\nab\cdot\delta \B =0,\\
\stq\label{dp_par}
&&\delta p_{||}=-2p_{||}\delta B_{||}/B,\\
\stq\label{dp_per}
&&\delta p_\perp=p_\perp\delta B_{||}/B,
\end{eqnarray}
where $\delta \v=\partial \xxi/\partial t$,
$\Xi=1+2(c^2_\perp-c^2_{||})/v^2_{\rm A}$, and we ignore the
perturbation in the plasma density, i.e. $\delta\rho\simeq 0$. Here
$\rho, p_{||}, p_\perp, \v,$ and $\B$ are the unperturbed quantities
while $\delta p_{||}, \delta p_\perp, \delta\v,$ and $\delta\B$ are
the perturbed quantities. Equations (\ref{dp_par}) and
(\ref{dp_per}) are calculated from the two equations of state for
$p_{||}$ and $p_\perp$ that are known as double adiabatic equations,
${d \over dt}(p_{||}B^2/\rho^3)=0$ and ${d\over dt}(p_\perp/(\rho
B))=0$ \cite{CGL56}. $\Xi=1$ if the pressure is isotropic, i.e
$p_{||}=p_\perp=p$.

To avoid complexities, we shall ignore the rotation of the star and
consequently neglect both the toroidal field $B_\phi$ and velocity
$v_\phi$. These assumptions simplify our calculations significantly,
and, we believe, allow the model to retain the important physics.
Furthermore, we do not consider a jump condition in the enhanced
density region as discussed by \cite{LN04}.  As a result, we do not
address Rayleigh-Taylor and/or Kelvin-Hemholtz instabilities.

%----------------------------------------------------------------------------------------

%\subsection{The dipolar magnetic field}

We expand the MHD equations (\ref{eq-mhd1}) in the orthogonal
coordinate system ($\mu, \nu, \phi$), where $\mu=\cos\theta/r^2$ is
the magnetic field-aligned coordinate, $\nu=\sin^2\theta/r$
numerates magnetic shells in the direction perpendicular to the
field line, and $\phi$ the is azimuthal coordinate. The components
of the metric are $h_\mu=h_\nu h_\phi$,
$h_\nu=r^2/\sin\theta\sqrt{1+3\cos^2\theta}$, and
$h_\phi=r\sin\theta$ where ($r,\theta,\phi$) is the spherical
coordinate. The metric component $h_\mu$ allows a convenient
representation of the dipolar magnetic field in the form
$\B=\mu^{\rm mag}/h_\mu \e_\mu$ where $\mu^{\rm mag}$ is the
magnetic dipole moment of the star.  Note that, for a dipolar
magnetic field $\B=B_p\e_\mu$, $B_p h_\mu=\mu^{\rm mag}$ is constant
which leads to $\nab\times\B=0$. We further assume that the ambient
flow velocity is along magnetic field lines, ie. $\v=v_p\e_\mu$.

Assuming \st \be \delta (\mu,\nu,\phi,t) \sim \delta(\nu) e^{i k\mu}
e^{-i\omega t} e^{im\phi}. \ee we can reduce the equations of motion
(\ref{eq-mhd1}) to one second order differential equation for
$\delta v_\nu$ as
\begin{widetext}
\st
\bea\label{diff-v-nu}
&&\frac{d^2\delta v_\nu}{d \nu^2}
+ F(\mu,\nu) \frac{d\delta v_\nu}{d \nu}
+ G(\mu,\nu) \delta v_\nu=0,\\
\stq
&&
F(\mu,\nu)=\frac{1}{(c^2_\perp + v_A^2)\eta_1}\ls (c^2_\perp+v^2_A)
\partial_\nu(\eta_1+ \eta_2) +
2\Xi {v^2_{\rm A}\over h^2_\mu}\eta_1\partial_\nu \ln h_\mu
+{2ik v_p  \eta_1 (3c^2_{||}-c^2_\perp) \partial_\nu \ln h_\mu\over
h_\mu (i\omega_D+K_\mu-\nab\cdot\v_p)}\rs
,\\
\stq
&&G(\mu,\nu)=\frac{1}{(c^2_\perp+v^2_A)\eta_1}
\ls (c^2_\perp+ v^2_A) \partial_\nu\eta_2
+2\Xi {v^2_A\over h^2_\mu}\eta_2\partial_\nu \ln h_\mu
- (i\omega_D-K_\nu)h_\nu
-{h_\nu \Xi v^2_A \over h^2_\mu}{k^2+(\partial_\mu \ln h_\nu)^2
\over i\omega_D+K_\nu-\nab\cdot v_p}
 \right. \no\\
&&\hspace{6cm} \left.
 +{2v_p\partial_\nu \ln h_\mu/h_\mu \over i\omega_D-K_\mu-\nab\cdot v_p}
\lp ik(3c^2_{||}-c^2_\perp)\eta_2 - \partial_\nu(h_\mu v_p)/h_\nu\rp
 \rs \,,\\
\stq
&&\eta_1=  (h_\phi Q/h_\mu)/[ (i\omega_D  - K_\mu)Q
- (m^2/h^2_\phi)(c^2_\perp+v^2_{\rm A})~ (i\omega_D + K_\phi - \nab\cdot\v_p)],\\
\stq
&&\eta_2=
(\eta_1/ h_\phi) \ls \frac{1}{h_\mu} \partial_\nu(h_\mu h_\phi)
- \frac{2}{h_\nu} \partial_\nu h_\mu  +{1\over h_\mu h_\nu}(h_\mu\partial_\nu v_p
 - v_p\partial_\nu h_\mu )
 ~{i k - \partial_\mu h_\nu /h_\nu\over
i\omega_D + K_\nu - \nab\cdot\v_p}\rs,\\
\stq\label{Q}
&&
Q=\omega_D^2 + K^2_\phi + (i\omega_D-K_\phi)\nab\cdot\v_p -
{v^2_A\Xi\over h^2_\mu} [ k^2 +\frac{1}{h^2_\phi}(\partial_\mu h_\phi)^2],
\eea
\end{widetext}
where $\omega_D=\omega-k v_p/h_\mu$ is the Doppler shifted
frequency, $K_i= \v_p\cdot\nab\ln h_i$ ($i=\mu, \nu, \phi$), $v_{\rm
A}=B/\sqrt{4\pi\rho}$ is the Alfv\'en wave velocity,
$c_{||}=\sqrt{p_{||}/\rho}$ and $c_\perp=\sqrt{p_\perp/\rho}$ are
sound velocities parallel and perpendicular to the direction of the
magnetic field. We note that in deriving Eqs. (\ref{diff-v-nu}) we
assumed that $\partial_\nu [p_\perp+B^2/(8\pi)]\simeq 0$.

The shear Alfv\'en resonance happens at $\eta_1=0$ or equivalently
at $Q=0$ leading to \st \be\label{omega_res} \omega_D^2 + i
\nab\cdot\v_p ~\omega_D  + K^2_\phi -K_\phi \nab\cdot\v_p - {v^2_A
\Xi\over h^2_\mu} [ k^2 +\frac{1}{h^2_\phi}(\partial_\mu
h_\phi)^2]=0\,. \ee As a result, resonance frequencies will be given
by
%\begin{widetext}
\st
\bea\label{omega_res1}
&&\omega_\pm= k v_p/h_\mu -(i/2) \nab\cdot\v_p \pm (1/ 2)
\sqrt{\Delta},\\
&&\Delta=
4v^2_A \Xi[k^2+(\partial_\mu\ln h_\phi)^2]/h^2_\mu-( \nab\cdot\v_p-2K_\phi)^2
.\no
\eea
%\end{widetext}
For a rectilinear configuration,  resonance eigenfrequencies Eq.
(\ref{omega_res1}) for an incompressible plasma flow along the field
lines, $\nab\cdot\v_p=0$, with an isotropic pressure,
$p_{||}=p_\perp=p$, reduce to ones we obtained in paper I, and as
expected: $\omega_\pm= k( v_p \pm v_A)$.

Equation (\ref{omega_res1}), however, shows that whenever the
ambient flow is compressible, i.e. $\nab\cdot\v_p\ne 0$, and/or
$\Delta <0$, waves do not propagate and an instability develops. In
general, the ambient plasma is fairly incompressible, i.e.
$\nab\cdot\v_p = 0$. However, due to the topological deformation of
the magnetosphere caused by the compressional action of accreting
material, a non-zero density gradient through the plasma, and so
$\nab\cdot\v_p\ne 0$ can be expected: \st \bea
&&\nab\cdot\v_p=-{1\over \rho} \frac{d\rho}{dt}=
-{1\over \rho}{\partial \rho \over \partial t} - {1\over \rho} \v_p\cdot\nab \rho,\no\\
&&\hspace{1.1cm}\simeq -  \v_p\cdot\nab\ln \rho .
\eea
As a result, the resonant mode will grow (decay) if
$\v_p\cdot\nab \rho>0$ ($<0$).
Approximating the plasma inflow velocity with the free fall
velocity $v_p\sim v_{\rm ff}(r)=(2GM/r)^{1/2}$, the growth/decay
timescale will be as order
of $\tau\sim r/v_p=(r^3/2GM)^{1/2}\sim 6\times 10^{-5}\;
(M/M_\odot)^{-1/2}(r/10\,{\rm km})^{3/2}$ s.
Therefore, the closer to the star the faster the
instability develops.

The condition $\Delta <0$ is satisfied whether
\st
\bea\label{Delta_eq}
\stq\label{firehose}
&&{\rm I}:~~~ \Xi <0 \rightarrow c^2_{||} > c^2_\perp+v^2_{\rm A}/2,\\
\stq\label{Delta1} &&{\rm II}:~~~ (\nab\cdot\v_p-2K_\phi)^2>
4v^2_A|\Xi|[k^2+(\partial_\mu\ln h_\phi)^2]/h^2_\mu . \no\\
\eea
Case I, that is known as the firehose instability in the literature,
happens when $p_{||}$ is much larger than $p_\perp +p_{\rm M}$,
where $p_{\rm M}=B^2/(8\pi)$ is the magnetic pressure.

The magnetic field channeling the parallel plasma streams
experiences a similar instability. Whenever the flux tube is
slightly bent, the flowing plasma exerts a centrifugal force, that
tends to enhance the initial bending. The field line bending is
proportional to the density of energy in plasma motion along the
magnetic field $\sim \rho v^2_{||}\sim p_{||}$. Recalling Eq.
(\ref{p_par}), the ambient plasma flow would enhance the firehose
instability in the magnetosphere of an accreting neutron star: \st
\be\label{p_par1} c^2_{||}\simeq {c'}^2_{||}  + v^2_p, \ee where
${c'}^2_{||}\simeq (m/\rho) \int f u^2_{||} d^3u$ and we assume that
a Maxwellian distribution function, so the cross term vanishes.
Inserting Eq. (\ref{p_par1}) into Eq. (\ref{firehose}), we find \st
\be\label{firehose1} v^2_p> v^2_{\rm A}/2+ ( c^2_\perp -{c'}^2_{||}
). \ee Therefore, a firehose instability develops whenever condition
(\ref{firehose1}) is satisfied.

For an isotropic pressure, i.e. $\Xi\sim 1$, the firehose
instability would not be expected.  However, an instability may
arise if the condition (\ref{Delta1}) is satisfied. For an
incompressible flow, we find that Eq. (\ref{Delta1}) reduces to
\st
\be\label{cond} v_p > v_{\rm A} \sqrt{1+q^2},
\ee
where $q=k/(\partial_\mu \ln h_\phi)$.
It is necessary to note that this
instability is enhanced whenever both $v_p\ne 0$ and $\partial_\mu
\ln h_\phi\ne 0$.  The latter condition is due to the curvature of
magnetic field lines.  Therefore, the non-flat topology of the
magnetic field can trigger some MHD instabilities through the
magnetosphere.  An interesting note is that the condition
(\ref{Delta1}) is only valid for for an isotropic pressure flow.
When the wave transfers energy to the flow, the wave decays, and the
extraction of flow energy by the wave will lead to a growing mode.
For a superAlfv\'enic flow, the extraction of energy from the flow
and growing MHD waves is very likely \cite{JNR97}.  In this case
also the growth/decay timescale will be as order of $\tau\sim
r/v_p=(r^3/2GM)^{1/2}\sim 6\times 10^{-5}\;
(M/M_\odot)^{-1/2}(r/10\,{\rm km})^{3/2}$ s.

Furthermore, by approximating the Alfv\'en velocity by $v_{\rm
A}\sim B(r)/\sqrt{4\pi \rho_{\rm ff}}$ where $\rho_{\rm
ff}=\dot{M}/(v_{\rm ff}~ 4\pi r^2)$ is the free fall mass density
($v_p\sim v_{\rm ff}(r)=\sqrt{2GM/r}$), the instability condition
(\ref{cond}) is satisfied for distance further than \st
\be\label{cond2} r>1.7\times 10^6\; {\rm cm}
\;(1+q^2)^{2/7}\mu^{4/7}_{26} \dot{M}_{17}^{-2/7}
(M/M_\odot)^{-1/7}. \ee Therefore, this instability is very likely
to develop at a position where the accretion disk can distort the
dipolar magnetosphere. Here $\mu_{26}$ is the magnetic field dipole
moment at the surface of star in units of $10^{26}$ G cm$^{3}$ and
$\dot{M}_{17}$ is the mass of accretion rate in units of $10^{17}$ g
s$^{-1}$.  We calculate Eq. (\ref{cond2}) for a solar mass neutron
star with 10 km radius.

To summarize, we believe we have shown that there are generic plasma
instabilities associated with the radial position where the
inflowing material of the accretion disk leads to a distortion of
the inner magnetosphere of the neutron star and field aligned plasma
flows. We suspect that these instabilities may be relevant in the
understanding of some details of observed QPOs in X-ray binary
systems, particularly when linked to shear Alfv\'en waves. The MHD
interaction of the infalling plasma with the neutron star's
magnetosphere can alter the topology of the star's inner
magnetoshere. The plasma flows and topology lead to the possibility
of : (1)  firehose instabilities associated the pressure anisotropy
produced by the plasma flow; (2) convective growth of waves in the
plasma flow.  The unstable modes might produce shear Alfv\'en
resonances with large amplitude giving strong quasi-periodic
variations in X-ray fluxes.

%-----------------------------------------------------------------------------------------

%----------------------------------------------------------------------------
%\subsubsection{A compressible flow}

\begin{acknowledgments}
This research was supported by the Natural Sciences and
Engineering Research Council of Canada (NSERC).
\end{acknowledgments}

%\newpage %Just because of unusual number of tables stacked at end
%\bibliography{apssamp}% Produces the bibliography via BibTeX.
%\end{document}
%--------------------------------------------------------------------------------------------
%
\def\aj{{AJ}}                   % Astronomical Journal
\def\araa{{ARA\&A\ }}             % Annual Review of Astron and Astrophys
\def\apj{{ApJ\ }}                 % Astrophysical Journal
\def\apjl{{ApJ\ }}                % Astrophysical Journal, Letters
\def\apjs{{ApJS\ }}               % Astrophysical Journal, Supplement
\def\apss{{Ap\&SS}}             % Astrophysics and Space Science
\def\aap{{A\&A\ }}                % Astronomy and Astrophysics
\def\aapr{{A\&A~Rev.}}          % Astronomy and Astrophysics Reviews
\def\aaps{{A\&AS}}              % Astronomy and Astrophysics, Supplement
\def\azh{{AZh}}                 % Astronomicheskii Zhurnal
\def\baas{{BAAS}}               % Bulletin of the AAS
\def\jrasc{{JRASC}}             % Journal of the RAS of Canada
\def\memras{{MmRAS}}            % Memoirs of the RAS
\def\mnras{{MNRAS\ }}             % Monthly Notices of the RAS
\def\pra{{Phys.~Rev.~A}}        % Physical Review A: General Physics
\def\prb{{Phys.~Rev.~B}}        % Physical Review B: Solid State
\def\prc{{Phys.~Rev.~C\ }}        % Physical Review C
\def\prd{{Phys.~Rev.~D\ }}        % Physical Review D
\def\pre{{Phys.~Rev.~E}}        % Physical Review E
\def\prl{{Phys.~Rev.~Lett.\ }}    % Physical Review Letters
\def\pasp{{PASP}}               % Publications of the ASP
\def\pasj{{PASJ\ }}               % Publications of the ASJ
\def\qjras{{QJRAS}}             % Quarterly Journal of the RAS
\def\skytel{{S\&T}}             % Sky and Telescope
\def\solphys{{Sol.~Phys.}}      % Solar Physics
\def\sovast{{Soviet~Ast.\ }}      % Soviet Astronomy
\def\ssr{{Space~Sci.~Rev.\ }}     % Space Science Reviews
\def\zap{{ZAp}}                 % Zeitschrift fuer Astrophysik
\def\nat{{Nature\ }}              % Nature
\def\iaucirc{{IAU~Circ. No.}}       % IAU Cirulars
\def\aplett{{Astrophys.~Lett.}} % Astrophysics Letters
\def\apspr{{Astrophys.~Space~Phys.~Res.}}
                % Astrophysics Space Physics Research
\def\bain{{Bull.~Astron.~Inst.~Netherlands}}
                % Bulletin Astronomical Institute of the Netherlands
\def\fcp{{Fund.~Cosmic~Phys.}}  % Fundamental Cosmic Physics
\def\gca{{Geochim.~Cosmochim.~Acta}}   % Geochimica Cosmochimica Acta
\def\grl{{Geophys.~Res.~Lett.}} % Geophysics Research Letters
\def\jcp{{J.~Chem.~Phys.}}      % Journal of Chemical Physics
\def\jgr{{J.~Geophys.~Res.}}    % Journal of Geophysics Research
\def\jqsrt{{J.~Quant.~Spec.~Radiat.~Transf.}}
                % Journal of Quantitative Spectroscopy and Radiative Trasfer
\def\memsai{{Mem.~Soc.~Astron.~Italiana}}
                % Mem. Societa Astronomica Italiana
\def\nphysa{{Nucl.~Phys.~A}}   % Nuclear Physics A
\def\nphysb{{Nucl.~Phys.~B\ }}   % Nuclear Physics A
\def\physrep{{Phys.~Rep.}}   % Physics Reports
\def\physscr{{Phys.~Scr}}   % Physica Scripta
\def\planss{{Planet.~Space~Sci.}}   % Planetary Space Science
\def\procspie{{Proc.~SPIE}}   % Proceedings of the SPIE

----------------------------------------------------------------------------------------

\end{document}